\documentclass[runningheads]{llncs}

\usepackage[utf8]{inputenc}
\usepackage[T1]{fontenc}
\usepackage{amsfonts}
\usepackage{amsmath}
\usepackage{hyperref}
\usepackage[capitalise]{cleveref}
\usepackage{placeins}
\usepackage{cite} 
\newcommand{\citep}{\cite} 
\usepackage{graphicx}
\usepackage{booktabs} 
\usepackage{adjustbox}
\usepackage[caption=false]{subfig}
\usepackage{setspace}

\usepackage{microtype}

\usepackage{contour}
\usepackage{ulem}
\contourlength{0.8pt}

\begin{document}

\title{Robust and efficient computation of retinal fractal dimension through deep approximation}

\titlerunning{Robust and efficient deep approximation of retinal fractal dimension}

\makeatletter
\newcommand{\printfnsymbol}[1]{%
  \textsuperscript{\@fnsymbol{#1}}%
}
\makeatother

\author{Justin Engelmann\inst{1} 
\and Ana Villaplana-Velasco\inst{2}
\and Amos Storkey\inst{3}\thanks{Equal supervision.}
\and \\Miguel O. Bernabeu\inst{2}\printfnsymbol{1}}

\institute{CDT Biomedical AI, School of Informatics, University of Edinburgh\\
\email{justin.engelmann@ed.ac.uk}
\and Centre for Medical Informatics, University of Edinburgh
\and School of Informatics, University of Edinburgh}

\maketitle 
\begin{abstract}
A retinal trait, or phenotype, summarises a specific aspect of a retinal image in a single number. This can then be used for further analyses, e.g. with statistical methods. However, reducing an aspect of a complex image to a single, meaningful number is challenging. Thus, methods for calculating retinal traits tend to be complex, multi-step pipelines that can only be applied to high quality images. This means that researchers often have to discard substantial portions of the available data. We hypothesise that such pipelines can be approximated with a single, simpler step that can be made robust to common quality issues. We propose Deep Approximation of Retinal Traits (DART) where a deep neural network is used predict the output of an existing pipeline on high quality images from synthetically degraded versions of these images. We demonstrate DART on retinal Fractal Dimension (FD) calculated by VAMPIRE, using retinal images from UK Biobank that previous work identified as high quality. Our method shows very high agreement with FD\textsuperscript{VAMPIRE} on unseen test images (Pearson $r=0.9572$). Even when those images are severely degraded, DART can still recover an FD estimate that shows good agreement with FD\textsuperscript{VAMPIRE} obtained from the original images (Pearson $r=0.8817$). This suggests that our method could enable researchers to discard fewer images in the future. Our method can compute FD for over 1,000img/s using a single GPU. We consider these to be very encouraging initial results and hope to develop this approach into a useful tool for retinal analysis.

\end{abstract}
\section{Introduction}
Retinal fundus images are non-invasive and low-cost. They are important for ophthalmology and also capture a detailed picture of the retinal vasculature. Thus, they can be used for studying and potentially predicting diseases such as diabetes, stroke, hypertension and neurovascular disease \citep{macgillivray2014retinal}. To analyse the relationships between aspects of the retina and other quantities of interest, retinal traits (also called  features, parameters or phenotypes) are used as a quantitative description of a specific aspect of the retinal image. Reducing a complex image to a single, meaningful number is necessary to use standard statistical methods yet a challenging task. It is challenging to identify a potentially salient aspect of the retina in the first place and to then design a method that can reliably quantify this aspect. This is further complicated by the large variability in retinal images stemming from idiosyncrasies of the imaged retinas (e.g. due to retinal diseases or rare phenotypes) and image quality (e.g. due to operator inexperience or time pressures in large scale cohort studies). Thus, pipelines for extracting such retinal traits tend to be complex and comprise of multiple steps, and can only be applied to images of sufficient quality. 

Poor image quality is a key problem in retinal image analysis. Particularly for large scale studies such as UK Biobank, many images are of poor quality being blurred, obscured, or hazy \citep{macgillivray2015suitability}. Imaging artefacts such as noise, non-uniform illumination or blur can also lead to poor vessel segmentations \citep{mookiah2021review}. Previous work analysing 2,690 UK Biobank participants found that only 60\% had an image that could be adequately analysed by VAMPIRE \citep{macgillivray2015suitability}. Two recent large-scale studies using retinal Fractal Dimension (FD) for predicting cardiovascular disease risk discarded 26\% \citep{zekavat2022deep} and 43\% \citep{velasco2021decreased} of the images in UK Biobank. Although necessary, this is unfortunate as it leads to lower sample sizes and makes it hard to study rare diseases in particular.

\begin{figure}[!t]
     \centering
    \includegraphics[width=0.95\textwidth]{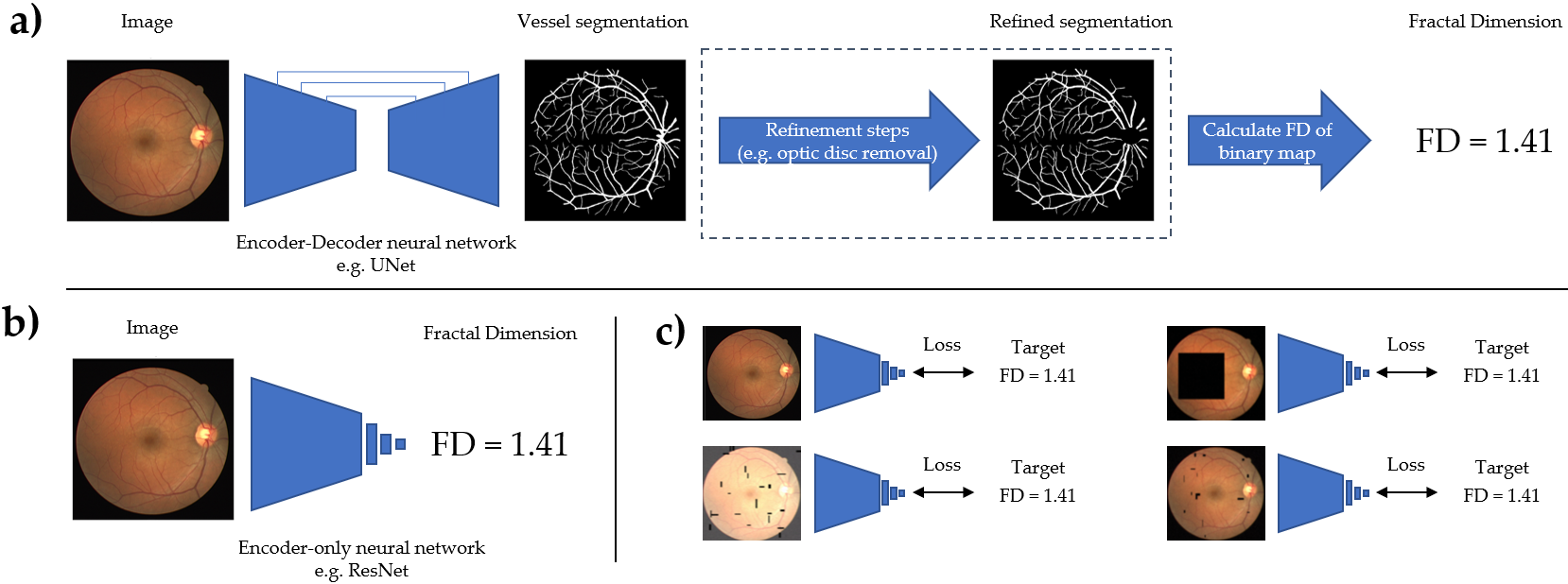}
    \caption{\textbf{Overview of our proposed framework.} a) A typical pipeline for computing FD: an encoder-decoder neural network for segmentation, potentially some refinement steps like optic disc segmentation and removal, and a method to calculate FD of the segmentation (e.g. box counting or multifractal). b) DART, our proposed approach outputs a deep approximation of FD in a single step using an encoder-only neural network, with drastically reduced complexity. c) We can train our model to be robust to image quality issues by synthetically degrading input images and training our model to minimise the loss between its output and the FD obtained with the original high quality image.}
    \label{fig:idea}
\end{figure}

We hypothesise that it is possible to approximate pipelines for calculating retinal traits with a single, simpler step and propose Deep Approximation of Retinal Traits (DART). \cref{fig:idea} gives a high-level overview of our approach. DART trains a deep neural network (DNN) to predict the output of an original method (OM) for calculating a retinal trait. We can then train the model to be robust to image quality issues by synthetically degrading the input images during training and asking the DNN model to predict the output of the OM on the original high quality image. The intuition behind this approach is that obtaining a high quality segmentation of the entire retina is a much harder task than describing an aspect of the vasculature like vascular complexity directly. DART offers a segmentation-free way of computing retinal traits related to the vasculature, but can also be applied to any other retinal image analysis method like feature extraction for disease grading or pathology segmentation. 

In the present work, we focus on retinal Fractal Dimension (FD), a key retinal trait that has been used to predict cardiovascular disease risk \citep{velasco2021decreased,zekavat2022deep} and is associated with neurodegeneration and stroke \citep{lemmens2020systematic}. We use FD as calculated by VAMPIRE \citep{trucco2013novel} with the multifractal \citep{stosic2006multifractal} method as the OM we apply DART to. At minimum, FD\textsuperscript{DART} should have very high agreement with FD\textsuperscript{VAMPIRE} on high quality images so that it can be interpreted in the same way. To be a useful method, it should further be robust to image quality issues and efficient. Robustness would enable researchers to discard fewer images than currently necessary while efficiency allows to conduct analyses at large scale without requiring large compute resources.

\section{Deep Approximation of Retinal Traits (DART)}
\subsection{Motivation and theory}
We hypothesise that it is possible to approximate the entire pipeline of an original method (OM) for calculating a retinal trait in a single, simpler step. We denote the distribution of high quality retinal fundus images as $X^{HQ}$, where each image $x_i$ has dimensions height H, width W, and channels C. The OM can be interpreted as a function $f$ that maps from the image space to one-dimensional retinal trait space (in our case, FD) $f:\mathbb{R}^{HxWxC}\to \mathbb{R}^1$, i.e. given an image $x_i\in X^{HQ}$ the FD computed by the OM is $FD^{OM}=f(x_i)$. Our goal is to find an alternative function $g:\mathbb{R}^{HxWxC}\to \mathbb{R}^1$ that is both simpler than $f$ and has high agreement with $f$ for all images of sufficient quality that the OM can be used, i.e. for all $x_i\in X^{HQ}$ $f(x_i)\approx g(x_i)$.

Designing such a simpler function by hand would be very challenging. Thus, we use a deep neural network (DNN). DNNs are universal function approximators in theory and very effective for image analysis in practice. We can then find a good approximation of $f$ by simply updating the model parameters $\theta$ (weights, biases, normalisation layer parameters) to minimise some differentiable measure of divergence between $f(x_i)$ and $g(x_i)$, e.g. mean squared error.

\subsubsection{Accuracy}
The output of the OM is fully determined by the given image, so we would expect that very high accuracy can be achieved. This contrasts with other problems, e.g. clincians take into account additional information like symptoms and family history, and might disagree with each other or even themselves if shown the same image multiple times.

\subsubsection{Simplicity \& Efficiency}
Some readers might not perceive DNNs as simple or efficient. However, modern pipelines for retinal image analysis tend to use DNNs for vessel segmentation, so not requiring additional steps implies strictly lower complexity both computationally and in terms of required code. Furthermore, segmentation models tend to have an encoder-decoder structure (e.g. UNet) whereas models for classification/regression only need an encoder and small prediction head, making them more parameter-, memory-, and compute-efficient. Finally, given the widespread adoption of deep learning, the frameworks are very mature and can be very efficiently GPU-accelerated.

\subsubsection{Robustness} We hypothesise that there images of lower quality that are such that a) current pipelines would not produce a useful FD number, but b) there is still sufficient information to give an accurate estimate of the FD number we would have obtained on a counterfactual high quality image. For example, in an image with an obstruction, only part of the retina might be visible. Thus, the resulting vessel segmentation map would be poor and the FD of this map would be very different from that of the counterfactual high quality image, yet the visible parts of the retina might contain sufficient information about the vascular complexity of the retina as a whole to recover an accurate estimate of the FD.

As we do not observe counterfactual high quality images or objective ground truth FD values, we artificially degrade high quality images with a degradation function $\mathrm{degrade}(x_i)=x_i^{\mathrm{degarded}}$ and train our model to minimise the difference between the predicted FD for the degraded image and the OM's FD for the high quality image $g_\theta(x_i^{\mathrm{degarded}})\approx f(x_i)$. If there indeed is sufficient information in the degraded images, then our model should be able to predict the OM's FD from the high quality image reasonably well. However, this is a much harder task than matching the OM on high quality images, as the degradations lose information and for a given degraded image there are multiple possible counterfactual high quality images.

\subsection{Implementation}

\subsubsection{Model \& Training}
Our model consists of a pretrained ResNet18 \citep{he2016deep} backbone that extracts a feature map from the images, followed by spatial average pool and a small multi-layer perceptron with a two hidden layers with 128 and 32 units, and a single output. Each hidden layer is followed by a layernorm \citep{ba2016layer} and GELU \citep{hendrycks2016gaussian} activation. No activation is applied to the final output. ResNet is a well-established architecture that has been shown to perform competitively with more recent architectures when using modern training techniques \citep{bello2021revisiting,wightman2021resnet}. We use Resnet18 as it is the most light-weight member of the Resnet family. We initialise the backbones with pre-trained weights on natural images from Instagram \citep{yalniz2019billion}. Those images are very different from retinal images, thus this is merely a minor refinement on random initialisation. We resize images to 224x224 pixels for computational efficiency and lower memory requirements. Apart from standard normalisation using channel-wise ImageNet mean and standard deviations, no further preprocessing is done and all 3 colour channels are kept. 

We train our model using a batchsize of 256 to minimise the mean squared error between prediction and target after normalizing the target to zero mean and unit variance, using mean and standard deviation from the training data to avoid data leakage. The model output can then be mapped back to FD range by applying the inverse transformation. We use the AdamW optimiser \citep{loshchilov2017decoupled} ($\beta_1 = 0.9, \beta_2 = 0.999$, weight decay of $10^{-6}$) and a cosine learning rate schedule \citep{loshchilov2016sgdr}. We train for 35 epochs with a linear learning rate warmup from $\eta_{min}=10^{-5}$ to $\eta_{max}=10^{-3}$ for 5 epochs, followed by 3 cycles of 10 epochs each. During each cycle, the current epoch learning rate is set according to a cosine schedule, and after each cycle $\eta_{max}$ is decayed by taking the square root. We apply generic data augmentations (horizontal ($p=0.5$) and vertical flip ($p=0.1$), mild affine transformations ($p=0.15$, rotation by up to ±10°, shear of up to ±5°, and scaling by ±5\%)) as well as the image degradations described in the next section with $p=0.75$ (sampling all 5 levels uniformly) to the images during training. We implemented our code in Python 3.9 using PyTorch and timm \citep{rw2019timm} and plan to make it publicly available upon publication.

\subsubsection{Synthetic degradations}
\begin{table}[!t]%
\caption{\textbf{Severity levels for the degradations.} Brightness, contrast and gamma changes are independently sampled from the given interval. Dimensions in pixels.}
\label{tab:synthetic_degradations_levels_params}
\centering
\begin{adjustbox}{max width=0.9\textwidth, max totalheight=1\textheight-2\baselineskip}
{\small

\begin{tabular}{@{}lc@{\hspace{0.5em}}c@{\hspace{0.5em}}c@{\hspace{0.5em}}c@{\hspace{0.5em}}c@{}}
\toprule
Severity                             & 1     & 2     & 3     & 4     & 5     \\ \midrule
Brightness/Contrast/Gamma            & ±5\%  & ±10\% & ±15\% & ±20\% & ±25\% \\
Mini Artifacts (holes, height, width) & 2-20/1-3/5-8 & 2-24/1-5/5-12 & 2-28/1-5/5-16 & 2-32/1-3/5-20 & 2-40/1-3/5-24 \\
Square Artifacts (side length)       & 25    & 50    & 75    & 100   & 125   \\
Chop Artifacts (\% of image removed) & 10-15 & 10-25 & 10-35 & 10-45 & 10-50 \\
Advanced Blur (kernel size, sigma)             & 3-5/0.2-0.5     & 3-7/0.2-0.7      & 3-9/0.2-0.8      & 3-11/0.2-0.9     & 3-13/0.2-1.0    \\
Gaussian Noise (variance)            & 1-10  & 5-10  & 5-20  & 5-25  & 5-30  \\ \bottomrule
\end{tabular}
}
\end{adjustbox}
\end{table}%

We focus on three types of quality issues in retinal images \citep{mookiah2021review,macgillivray2015suitability}: Lighting issues, artifacts/obstructions, and imaging issues. To simulate general lighting issues, we independently change brightness, contrast and gamma of the image. To simulate  artifacts/obstructions and severely inconsistent lighting, we introduce one of three artifacts: 1) many smaller rectangular holes placed across the retina, b) a single large square hole, or c) we ``chop'' off the bottom or top part of the image. The latter is inspired by the observation that in UK Biobank some images only have the top or bottom part properly illuminated. To simulate general imaging issues, we add pixel-wise Gaussian noise and blur the image. Standard isotropic Gaussian blur kernels do not mimic realistic image blur, so we use an advanced anisotropic blurring technique developed for image super-resolution \citep{wang2021real} where the standard deviations for both dimensions of the kernel are sampled independently, and the kernel is then rotated and has some noise added before being applied to the image.

We specify degradation parameters for five levels of severity, shown in \cref{tab:synthetic_degradations_levels_params}. For a given level, we sample parameters for each image independently from the given ranges. Degradations are applied after images have already been downsized to 224x224. We apply an artifact with $p=0.2*s$ where $s$ is the severity. If an image was chosen to have an artifact applied to it, we then choose Mini Artifacts with $p=0.85$, Square Artifact with $p=0.10$, and Chop Artifact with $p=0.05$. Degradations are implemented using the albumentations package \citep{info11020125}.

\begin{figure}[!th]
     \centering
    \includegraphics[width=\textwidth]{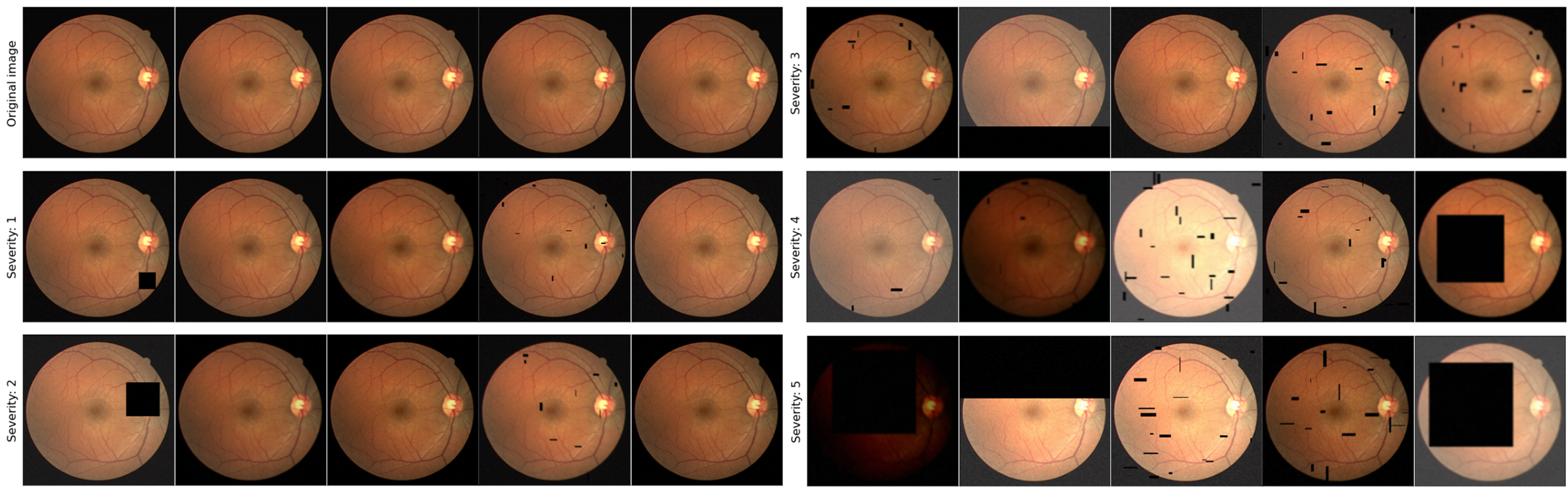}
    \caption{Random examples of synthetically degraded versions of the same fundus image. Best viewed zoomed in, especially for the advanced blur. UK Biobank asks to only reproduce imaging data where necessary, so we demonstrate the degradations on an image taken from DRIVE \citep{staal2004ridge} which is similar in appearance to those in UK Biobank.}
    \label{fig:degradations}
\end{figure}

\section{Experiments}
\subsection{Data}
We apply our DART framework multi-fractal FD  \citep{stosic2006multifractal} calculated with VAMPIRE \citep{trucco2013novel}. We use only images that had been identified as high quality in a previous study \citep{velasco2021decreased} as for those images FD\textsuperscript{VAMPIRE} should be reliable and can be considered as a reasonable ``ground-truth''. We randomly split the data into train, validation, and test sets containing 70, 10, and 20\% of the participants in UK Biobank, resulting in 52,242 / 7,478 / 14,907 images belonging to 32,300 / 4,614 / 9,229 participants in each set. We split at the participant level such that no images of the same participant occur in different sets. Images are cropped to square to remove black non-retinal regions and processed at 224x224 as described above.

\subsection{Results}
\subsubsection{Agreement \& Robustness}

\begin{table}[!t]%
\caption{Agreement between FD\textsuperscript{VAMPIRE} obtained on high quality images, and FD\textsuperscript{DART} for different levels of degradation measured on 14,907 held-out test set images.}
\label{tab:resultsvsserverity}
\centering
\begin{adjustbox}{max width=0.9\textwidth, max totalheight=1\textheight-2\baselineskip}
{\small
\begin{tabular}{l@{\hspace{1.4em}}l@{\hspace{1.4em}}l@{\hspace{1.4em}}l@{\hspace{1.4em}}l}
\toprule
Degradations & $R^2$ & Pearson $r$ (p-value) & Spearman $r$ (p-value) & OLS Regression fit \\
\midrule
None & 0.9160 & 0.9572 (0.0000) & 0.9561 (0.0000) & y=0.01 + 1.00x \\
 Severity 1 & 0.8957 & 0.9467 (0.0000) & 0.9446 (0.0000) & y=0.01 + 0.99x \\
 Severity 2 & 0.8859 & 0.9414 (0.0000) & 0.9396 (0.0000) & y=0.01 + 0.99x \\
 Severity 3 & 0.8623 & 0.9287 (0.0000) & 0.9282 (0.0000) & y=0.00 + 1.00x \\
 Severity 4 & 0.8309 & 0.9116 (0.0000) & 0.9103 (0.0000) & y=0.01 + 0.99x \\
 Severity 5 & 0.7773 & 0.8817 (0.0000) & 0.8840 (0.0000) & y=0.02 + 0.99x \\
\bottomrule
\end{tabular}

}
\end{adjustbox}
\end{table}%

We find very high agreement between FD\textsuperscript{VAMPIRE} and FD\textsuperscript{DART} on the original images with Pearson $r=0.9572$ and $r^2=0.9160$. \cref{tab:resultsvsserverity} shows results for different levels of degradations. When degrading the images and asking our model to predict the FD\textsuperscript{VAMPIRE} obtained from the high quality image, agreements goes down as the images become more degraded, which is what we would expect as these degradations remove substantial information about the retinal vasculature. However, despite this, we still observe good agreement with the FD\textsuperscript{VAMPIRE} obtained on the original image even at severity level 5 where extreme degradations are applied (Pearson $r=0.8817$ and $R^2=0.7773$). This suggests that DART can recover good estimates of the retinal trait that would have been obtained from a counterfactual high quality image even if the available image has very poor quality. Thus, this might allow for discarding much fewer images than currently necessary.

\begin{figure}[t]
     \centering
\subfloat[Scatterplots of FD\textsuperscript{DART} against FD\textsuperscript{VAMPIRE} obtained\\from original images for different levels of degradation.\label{fig:scatterplotresults}]{%
  \includegraphics[width=0.75\textwidth]{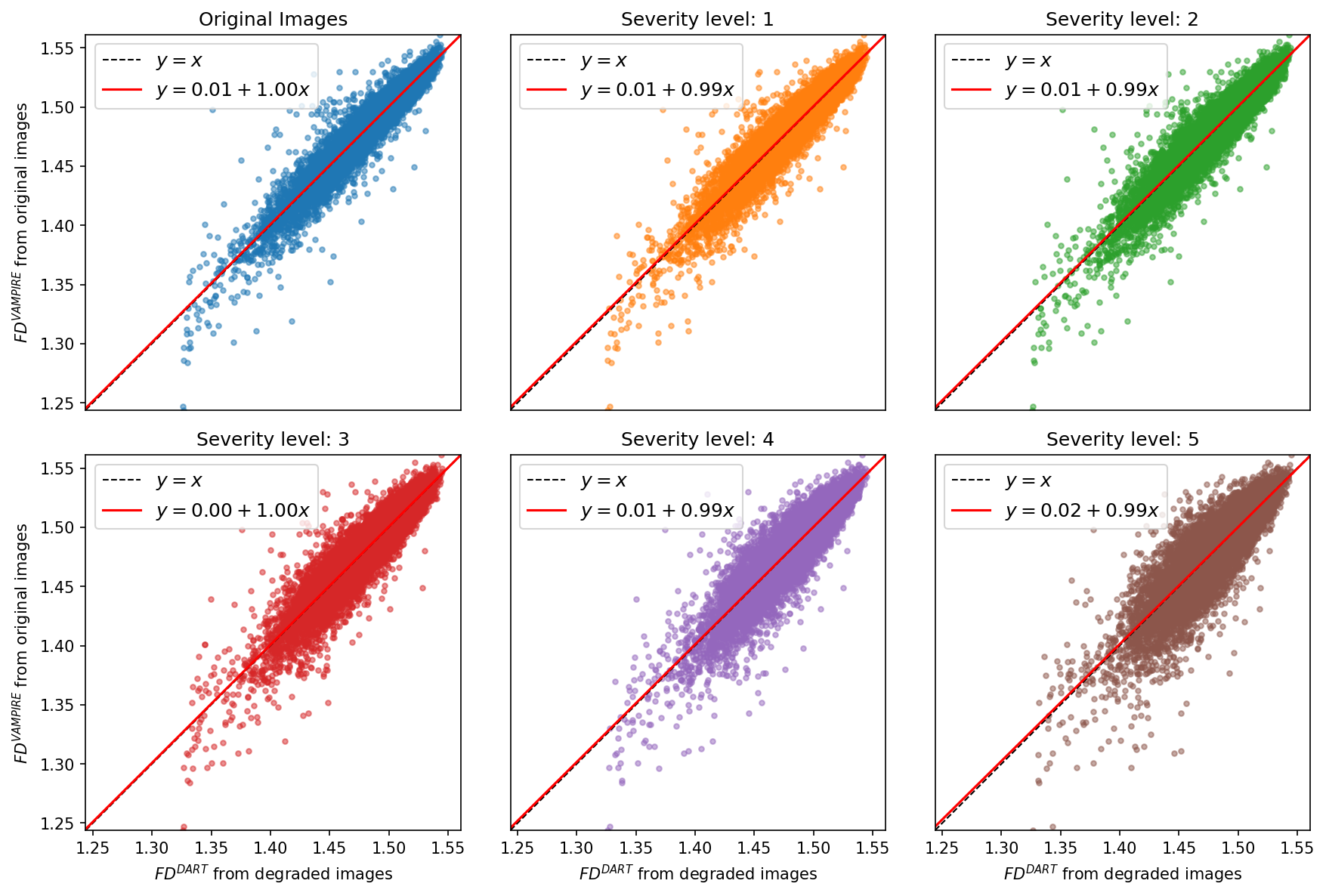}%
}\hfil
\subfloat[Boxplots of the\\ residuals.\label{fig:residualsresults}]{%
  \includegraphics[width=0.25\textwidth]{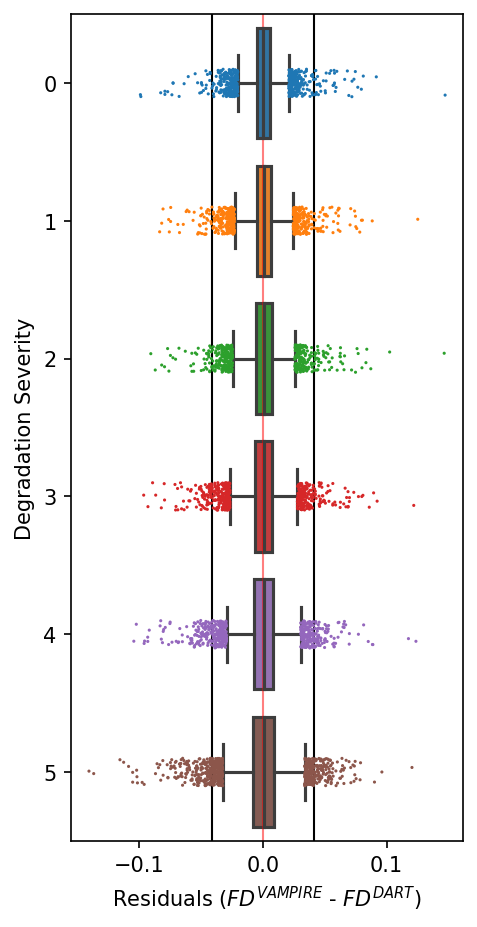}%
}
\caption{Agreement results for 14,907 held-out test set images. Best viewed zoomed in. \textbf{a)} Red line: best linear fit; dashed black line: $y=x$. \textbf{b)} Faint red line: $x=0$; vertical black lines: ± one interquartile range (IQR) of FD\textsuperscript{VAMPIRE} for reference.}
\label{fig:agreementresults}
\end{figure}

For comparison, a previous study comparing FD for arteries and veins separately between VAMPIRE and SIVA \citep{mcgrory2018towards} found very poor agreement between the measures of the two tools ($R^2=0.139$ and $R^2=0.168$ for arteries and veins, respectively). Another study comparing vessel caliber-related retinal traits obtained with VAMPIRE, SIVA, and IVAN found that they agreed with Pearson $r$s of 0.29 to 0.86. Thus, the observed agreement between FD\textsuperscript{VAMPIRE} and FD\textsuperscript{DART} with a Pearson $r=0.9572$ and $r^2=0.9160$ is very high, and even when DART is applied the most degraded images the agreement (Pearson $r=0.8817$ and $R^2=0.7773$) is higher than what could be expected when using two different tools on the same high quality images.

Finally, our method shows very low bias even as degradation severity is increased (\cref{fig:agreementresults}). The best OLS fit is very close to the identity line for all levels of severity, or equivalently, the optimal linear translation function from FD\textsuperscript{DART} to FD\textsuperscript{VAMPIRE} is almost simply the identity function. This also implies that no post-hoc adjustment for image quality is needed and FD\textsuperscript{DART} values obtained for images of varying quality are on the same scale out-of-the-box. As degradation severity increases, the variance of the residuals also increases but most residuals are still less than one interquartile range (IQR), a robust equivalent of the standard deviation, even when applying the strongest degradation.

\subsubsection{Speed}

Images were loaded into RAM so that hard disk speed is not a factor. We then measured the time it took to process all 52,242 training images, including normalisation, moving them from RAM to GPU VRAM, as well as the time to move the results back to RAM. We used a modern workstation (Intel i9-9920X 24 core CPU, single Nvidia RTX A6000 24GB GPU, 126GB of RAM) and a batchsize of 440. With ResNet18 as backbone, our model processed all 52,242 images in 48.5s ± 93.6 ms (mean±std over 5 runs), yielding a rate of 1077 img/s. 

\section{Conclusion}
We have shown that we can use DART to approximate the multi-step pipeline for obtaining FD\textsuperscript{VAMPIRE} with very high agreement. Our resulting model can compute FD\textsuperscript{DART} for over 1,000img/s using a GPU. Furthermore, our model can compute FD\textsuperscript{DART} values from severely degraded images that still match the FD\textsuperscript{VAMPIRE} values obtained on the high quality images well. This could allow researchers interested in studying retinal traits to discard fewer images than currently necessary and thus have higher sample sizes. We consider these to be very encouraging initial results.

There are a number of directions for future work. First, the proposed framework can be easily applied to other retinal traits like vessel tortuosity or width, or FD as calculated by other pipelines. We would expect that this would be similarly successful. Second, the robustness of the resulting DART model should be evaluated in more depth and the cases with extreme residuals should be manually examined. We expect that robustness can be further improved, especially if we identify common failure cases and use those as data augmentations. Third, many straight-forward, incremental technical improvements should be possible such as improved training procedures to further increase performance, trying different architectures and resolutions, and speeding up inference speed further through common tricks like fusing batch norm layers into the convolutional layers. Finally, we hope that our approach will eventually enable other researchers to conduct better analyses, e.g. by not having to discard as many images and thus having a larger sample size available.

\section*{Acknowledgements}
We thank our colleagues for their help and support.

This research has been conducted using the UK Biobank Resource under project 72144. This work was supported by the United Kingdom Research and Innovation (grant EP/S02431X/1), UKRI Centre for Doctoral Training in Biomedical AI at the University of Edinburgh, School of Informatics. For the purpose of open access, the author has applied a creative commons attribution (CC BY) licence to any author accepted manuscript version arising.

\bibliographystyle{splncs04}
\bibliography{references.bib}

\begin{thebibliography}{10}
\providecommand{\url}[1]{\texttt{#1}}
\providecommand{\urlprefix}{URL }
\providecommand{\doi}[1]{https://doi.org/#1}

\bibitem{ba2016layer}
Ba, J.L., Kiros, J.R., Hinton, G.E.: Layer normalization. arXiv preprint
  arXiv:1607.06450  (2016)

\bibitem{bello2021revisiting}
Bello, I., Fedus, W., Du, X., Cubuk, E.D., Srinivas, A., Lin, T.Y., Shlens, J.,
  Zoph, B.: Revisiting resnets: Improved training and scaling strategies. arXiv
  preprint arXiv:2103.07579  (2021)

\bibitem{info11020125}
Buslaev, A., Iglovikov, V.I., Khvedchenya, E., Parinov, A., Druzhinin, M.,
  Kalinin, A.A.: Albumentations: Fast and flexible image augmentations.
  Information  \textbf{11}(2) (2020). \doi{10.3390/info11020125},
  \url{https://www.mdpi.com/2078-2489/11/2/125}

\bibitem{he2016deep}
He, K., Zhang, X., Ren, S., Sun, J.: Deep residual learning for image
  recognition. In: Proceedings of the IEEE Conference on Computer Vision and
  Pattern Recognition. pp. 770--778 (2016)

\bibitem{hendrycks2016gaussian}
Hendrycks, D., Gimpel, K.: Gaussian error linear units (gelus). arXiv preprint
  arXiv:1606.08415  (2016)

\bibitem{lemmens2020systematic}
Lemmens, S., Devulder, A., Van~Keer, K., Bierkens, J., De~Boever, P., Stalmans,
  I.: Systematic review on fractal dimension of the retinal vasculature in
  neurodegeneration and stroke: assessment of a potential biomarker. Frontiers
  in neuroscience  \textbf{14}, ~16 (2020)

\bibitem{loshchilov2016sgdr}
Loshchilov, I., Hutter, F.: {SGDR: Stochastic gradient descent with warm
  restarts}. arXiv preprint arXiv:1608.03983  (2016)

\bibitem{loshchilov2017decoupled}
Loshchilov, I., Hutter, F.: Decoupled weight decay regularization. arXiv
  preprint arXiv:1711.05101  (2017)

\bibitem{macgillivray2015suitability}
MacGillivray, T.J., Cameron, J.R., Zhang, Q., El-Medany, A., Mulholland, C.,
  Sheng, Z., Dhillon, B., Doubal, F.N., Foster, P.J., Trucco, E., et~al.:
  Suitability of uk biobank retinal images for automatic analysis of
  morphometric properties of the vasculature. PLoS One  \textbf{10}(5),
  e0127914 (2015)

\bibitem{macgillivray2014retinal}
MacGillivray, T., Trucco, E., Cameron, J., Dhillon, B., Houston, J., Van~Beek,
  E.: Retinal imaging as a source of biomarkers for diagnosis, characterization
  and prognosis of chronic illness or long-term conditions. The British journal
  of radiology  \textbf{87}(1040),  20130832 (2014)

\bibitem{mcgrory2018towards}
McGrory, S., Taylor, A.M., Pellegrini, E., Ballerini, L., Kirin, M., Doubal,
  F.N., Wardlaw, J.M., Doney, A.S., Dhillon, B., Starr, J.M., et~al.: Towards
  standardization of quantitative retinal vascular parameters: comparison of
  siva and vampire measurements in the lothian birth cohort 1936. Translational
  vision science \& technology  \textbf{7}(2),  12--12 (2018)

\bibitem{mookiah2021review}
Mookiah, M.R.K., Hogg, S., MacGillivray, T.J., Prathiba, V., Pradeepa, R.,
  Mohan, V., Anjana, R.M., Doney, A.S., Palmer, C.N., Trucco, E.: A review of
  machine learning methods for retinal blood vessel segmentation and
  artery/vein classification. Medical Image Analysis  \textbf{68},  101905
  (2021)

\bibitem{staal2004ridge}
Staal, J., Abr{\`a}moff, M.D., Niemeijer, M., Viergever, M.A., Van~Ginneken,
  B.: Ridge-based vessel segmentation in color images of the retina. IEEE
  transactions on medical imaging  \textbf{23}(4),  501--509 (2004)

\bibitem{stosic2006multifractal}
Stosic, T., Stosic, B.D.: Multifractal analysis of human retinal vessels. IEEE
  transactions on medical imaging  \textbf{25}(8),  1101--1107 (2006)

\bibitem{trucco2013novel}
Trucco, E., Ballerini, L., Relan, D., Giachetti, A., MacGillivray, T., Zutis,
  K., Lupascu, C., Tegolo, D., Pellegrini, E., Robertson, G., et~al.: Novel
  vampire algorithms for quantitative analysis of the retinal vasculature. In:
  2013 ISSNIP Biosignals and Biorobotics Conference: Biosignals and Robotics
  for Better and Safer Living (BRC). pp.~1--4. IEEE (2013)

\bibitem{velasco2021decreased}
Velasco, A.V., Engelmann, J., Rawlik, K., Canela-Xandri, O., Tochel, C.,
  Lona-Durazo, F., Mookiah, M.R.K., Doney, A., Parra, E., Trucco, E., et~al.:
  Decreased retinal vascular complexity is an early biomarker of mi supported
  by a shared genetic control. medRxiv  (2021)

\bibitem{wang2021real}
Wang, X., Xie, L., Dong, C., Shan, Y.: {Real-ESRGAN}: Training real-world blind
  super-resolution with pure synthetic data. In: Proceedings of the IEEE/CVF
  International Conference on Computer Vision. pp. 1905--1914 (2021)

\bibitem{rw2019timm}
Wightman, R.: {PyTorch Image Models}.
  \url{https://github.com/rwightman/pytorch-image-models} (2019).
  \doi{10.5281/zenodo.4414861}

\bibitem{wightman2021resnet}
Wightman, R., Touvron, H., J{\'e}gou, H.: Resnet strikes back: An improved
  training procedure in timm. arXiv preprint arXiv:2110.00476  (2021)

\bibitem{yalniz2019billion}
Yalniz, I.Z., J{\'e}gou, H., Chen, K., Paluri, M., Mahajan, D.: Billion-scale
  semi-supervised learning for image classification. arXiv preprint
  arXiv:1905.00546  (2019)

\bibitem{zekavat2022deep}
Zekavat, S.M., Raghu, V.K., Trinder, M., Ye, Y., Koyama, S., Honigberg, M.C.,
  Yu, Z., Pampana, A., Urbut, S., Haidermota, S., et~al.: Deep learning of the
  retina enables phenome-and genome-wide analyses of the microvasculature.
  Circulation  \textbf{145}(2),  134--150 (2022)

\end{thebibliography}

\end{document}